\documentclass[letterpaper, 10 pt, conference]{ieeeconf}
\usepackage{silence}
\IEEEoverridecommandlockouts
\usepackage[T1]{fontenc}
\usepackage[latin1]{inputenc}
\usepackage{balance}
\usepackage{silence}
\usepackage{cite}
\usepackage{graphicx}
\usepackage{picinpar}
\usepackage{stfloats}
\usepackage{url}
\usepackage{flushend}
\usepackage{colortbl}
\usepackage{soul}
\usepackage{multirow}
\usepackage{pifont}
\usepackage{alltt}
\usepackage{enumerate}
\usepackage{siunitx}
\usepackage{pbox}

\usepackage{amsmath,amssymb,amsfonts}
\usepackage{amsthm}
\usepackage{algorithmic}
\usepackage{algorithm}
\usepackage{bm}
\usepackage{graphbox}
\usepackage{textcomp}
\usepackage{xcolor}
\usepackage[export]{adjustbox}
\usepackage{color}
\usepackage{subcaption}
\usepackage{booktabs}
\usepackage{microtype}
\def\BibTeX{{\rm B\kern-.05em{\sc i\kern-.025em b}\kern-.08em
    T\kern-.1667em\lower.7ex\hbox{E}\kern-.125emX}}

\usepackage{tikz}
\def\BibTeX{{\rm B\kern-.05em{\sc i\kern-.025em b}\kern-.08em
    T\kern-.1667em\lower.7ex\hbox{E}\kern-.125emX}}

\definecolor{LightCyan}{rgb}{0.8,0.8,1.0}
\definecolor{LightRed}{rgb}{1.0,0.8,0.8}
\definecolor{LightGreen}{rgb}{0.8,1.0,0.8}
\definecolor{LightYellow}{rgb}{1.0,1.0,0.8}

\newtheorem{theorem}{Theorem}

\ActivateWarningFilters[balance]

\makeatletter
\let\NAT@parse\undefined
\makeatother
\usepackage[hidelinks]{hyperref}

\newcommand{\calG}{{\cal G}}

\newcommand{\calN}{{\cal N}}

\newcommand{\bfk}{\mathbf{k}}

\newcommand{\bfq}{\mathbf{q}}

\newcommand{\bfu}{\mathbf{u}}

\newcommand{\bfx}{\mathbf{x}}

\newcommand{\bfalpha}{{\boldsymbol{\alpha}}}

\newcommand{\bftheta}{{\boldsymbol{\theta}}}

\title{\LARGE\bf Policy Gradient with Self-Attention for \\ Model-Free Distributed Nonlinear Multi-Agent Games}

\author{Eduardo Sebasti\'{a}n$^1$ \\ Eduardo Montijano$^3$ \and Maitrayee Keskar$^2$ \\ Carlos Sag\"{u}\'{e}s$^3$ \and Eeman Iqbal$^2$ \\ Nikolay Atanasov$^2$ %
\thanks{$^1$Department of Computer Science and Technology, University of Cambridge, UK (e-mail: \texttt{\small es2121@cam.ac.uk}).}
\thanks{$^2$Department of Electrical and Computer Engineering, University of California San Diego, USA (e-mails: \texttt{\small \{mmkeskar, eiqbal, natanasov\}@ucsd.edu}).}
\thanks{$^3$RoPeRt group, at DIIS - I3A, Universidad de Zaragoza, Spain (e-mails: \texttt{\small \{emonti, csagues\}@unizar.es}).}%
\thanks{This work has been supported by ONR N00014-23-1-2353, ONR Global grant N62909-24-1-2081, NSF CCF-2402689 (ExpandAI), in part by Spanish projects PID2024-159284NB-I00, PID2024-159279OB-I00, PID2021-124137OB-I00, by ERDF A way of making Europe and by the European Union NextGenerationEU/PRTR, via project REMAIN S1/1.1/E0111 (Interreg Sudoe Programme, ERDF), by DGA T45-23R, by a Spanish grant FPU19-05700 and by a US-Spain Fulbright grant.}%
}

\newcommand\copyrighttext{%
  \footnotesize \textcopyright This paper has been accepted for publication at IEEE/RSJ International Conference on Intelligent Robots and Systems. Please, when citing the paper, refer to the official manuscript.}
\newcommand\copyrightnotice{%
\begin{tikzpicture}[remember picture,overlay]
\node[anchor=south,yshift=10pt] at (current page.south) {\fbox{\parbox{\dimexpr\textwidth-\fboxsep-\fboxrule\relax}{\copyrighttext}}};
\end{tikzpicture}%
}

\begin{document}
\maketitle
\copyrightnotice


\begin{abstract}
Multi-agent games in dynamic nonlinear settings are challenging due to the time-varying interactions among the agents and the non-stationarity of the (potential) Nash equilibria. In this paper we consider model-free games, where agent transitions and costs are observed without knowledge of the transition and cost functions that generate them. We propose a policy gradient approach to learn distributed policies that follow the communication structure in multi-team games, with multiple agents per team. Our formulation is inspired by the structure of distributed policies in linear quadratic games, which take the form of time-varying linear feedback gains. In the nonlinear case, we model the policies as nonlinear feedback gains, parameterized by self-attention layers to account for the time-varying multi-agent communication topology. We demonstrate that our distributed policy gradient approach achieves strong performance in several settings, including distributed linear and nonlinear regulation, and simulated and real multi-robot pursuit-and-evasion games.
\end{abstract}


\section{Introduction}\label{sec:intro}

\begin{figure}
    \centering
    \includegraphics[width=1\columnwidth]{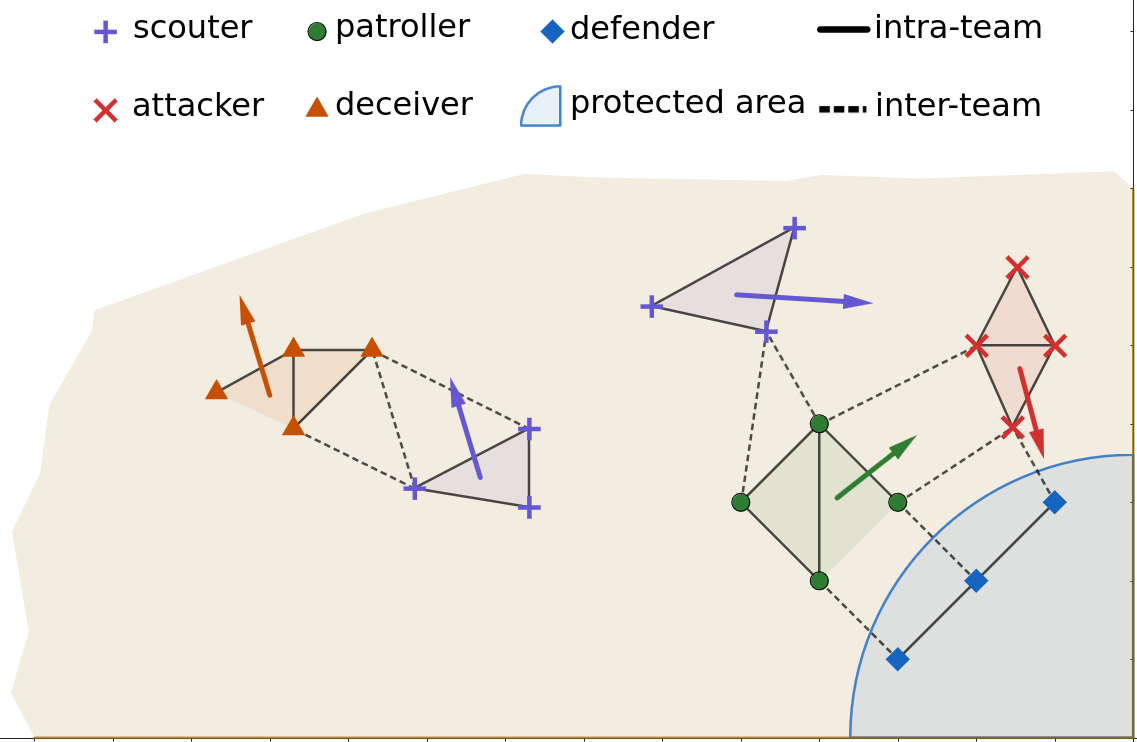}
    \caption{\small Multi-team perimeter-defense game. Each team has a different goal, number of agents, and constraints (e.g., the scouter team inspects the region to detect attackers whereas the deceiver team aims at confusing the scouters), leading to global cooperative and competitive behaviors (e.g., patrollers and defenders help each other to prevent attackers from crossing the perimeter). The agents only have access to local information, either from teammates (intra-team interactions) or agents on other teams (inter-team interactions). In general, the problem is nonlinear and dynamic with time-varying interactions and unknown transition and cost model.}
    \label{fig:perimeter_defense_example}
\end{figure}

Multi-robot problems encompass a variety of expected behaviors \cite{spica2020real,park2021multi,sebastian2022adaptive, sebastian2023physics}, including cooperative, conflicting or competitive actions. For instance, in a perimeter-defense setting \cite{shishika2020review, chen2023accelerated}, multiple teams must coordinate to effectively defend a region from potential attackers (Fig.~\ref{fig:perimeter_defense_example}). These kinds of problems can be formulated as multi-team dynamic games \cite{ parise2021analysis, lian2024nash}, where each multi-agent team is viewed as a player with specific goals and constraints, and where agents interact with teammates (intra-team interactions) and agents on other teams (inter-team interactions). These settings are typically nonlinear and dynamic, requiring complex interactions that evolve with time as a function of how the agents play the game. These features are specially relevant when we seek distributed policies subject to the communication constraints imposed by the teams' topology; and in the absence of a mathematical description of the game dynamics and costs, demanding model-free approaches that only rely on transition and cost samples. Inspired by solutions in linear quadratic games, we present a novel distributed policy structure for nonlinear dynamic games learned via policy gradients.

\textbf{Related Work}. Model-based methods for nonlinear multi-agent dynamic games rely on iterative linearization of the system dynamics and quadratic approximation of the game cost \cite{fridovich2020efficient}. This allows for fast computation with guarantees of convergence to a saddle configuration \cite{zheng2022escaping} but imposes a centralized calculation that limits the applicability in distributed settings, where the agent communication is restricted according to a graph topology. To overcome such limitations, it is possible to restrict the class of nonlinear multi-agent dynamic games to potential games \cite{williams2024potential, bhatt2025strategic}, where it is assumed that a potential function exists such that the relative incentives in modifying one agent's policy is equal to the difference in value of the potential function. Under this constraint, algorithms can be derived that compute open-loop optimal trajectories for the agents under centralized \cite{kavuncu2021potential} or distributed topological constraints \cite{ williams2024potential}. However, open-loop policies lack robustness and require knowledge on how the multi-agent topology will evolve with time. In contrast, we propose a novel policy parameterization that is distributed by construction and does not require network topology prediction. In all previous cases, a model of the dynamics and the cost function is needed to compute the actions. Instead, to address general nonlinear multi-agent dynamic games, we propose a model-free policy gradient approach that relies only on transition and cost samples.

Model-free solutions for games are limited due to the non-stationarity of the Nash equilibria (if one exists) \cite{zhang2024equilibrium}. Traditional approaches either focus on providing theoretical guarantees of convergence or addressing practical settings assuming the existence of such Nash equilibria. An instance of the former is \cite{furieri2020learning}, where distributed linear quadratic regulators are learned assuming that the sequence of graphs representing the communication structure of the game is known. From a different perspective, when the linear cost function is known and the strategies of all players are available, the problem can be posed as a multi-team distributed optimization program \cite{nguyen2024constrained, carnevale2024unifying}. In practical settings, existing solutions rely on multi-agent reinforcement learning algorithms \cite{yang2020overview} that consider independent heterogeneous agents to apply policy gradient methods \cite{giannou2022convergence, aydin2023policy}. In this work, we bring together the benefits of both alternatives by proposing a self-attention-based policy parameterization built from first principles and which enforces distributed execution constraints. The distributed policy is trained using a policy gradient learning method that considers, simultaneously, the policies of all the agents from all different teams, addressing the non-stationarity of the game in practice.

Closer to our setup, attention mechanisms have been combined with policy gradients to aggregate neighbor observations in cooperative multi-robot navigation \cite{huang2024collision}. We differ in two key respects: (i) we address multi-team dynamic games with both intra- and inter-team (competitive) interactions and heterogeneous roles, rather than a single cooperative swarm; and (ii) we do not use attention as a generic observation encoder, but to parameterize a nonlinear feedback gain derived from the structure of optimal distributed LQR policies.

\textbf{Contributions}. Our main contribution is a method for learning distributed policies in model-free nonlinear multi-agent dynamic games (Sec. \ref{sec:prosta}). Our approach uses a nonlinear feedback gain formulation of the agent policies, parameterized using self-attention layers (Sec. \ref{sec:distributed_parameterization}). The use of self-attention enables to enforce intra- and inter-team graph constraints, handling time-varying communication and achieving invariance with respect to the total number of agents. Furthermore, a neural network parameterization of the policies motivates the use of a policy gradient method to learn in the model-free setting.
The method also allows to learn heterogeneous policies per team, such that the teams adjust to specific goals. We demonstrate that our method applies broadly, from linear quadratic settings under topology constraints to multi-agent reinforcement learning in competitive games with simulated and real robots (Sec. \ref{sec:simulations}).


\section{Problem Statement}\label{sec:prosta}

We consider a discrete-time game among $\mathsf{N}$ multi-agent teams. Each team has $\mathsf{M}_i$ agents, where index $i$ refers to a specific team. The total number of agents is $\mathsf{M} = \sum_{i=1}^{\mathsf{N}}\mathsf{M}_i$. The agents interact in a distributed manner, described by a time-varying directed graph \mbox{$\mathcal{G}(k) = (\mathcal{V}, \mathcal{E}(k))$}. In graph $\calG(k)$, $k$ refers to the discrete time step and $\mathcal{V}$ is the set of agents, such that \mbox{$|\mathcal{V}| = \mathsf{M}$}. We denote by $\mathcal{V}_i$ the set of agents of team $i$, such that $|\mathcal{V}_i| = \mathsf{M}_i$, $\mathcal{V} = \bigcup_{i=1}^{\mathsf{N}}\mathcal{V}_i$ and \mbox{$\bigcap_{i=1}^{\mathsf{N}}\mathcal{V}_i = \emptyset$}. The set of edges is $\mathcal{E}(k)$ and $(l, p) \in \mathcal{E}(k)$ if and only if the information of agent $p$ is available to agent $l$ at instant $k$. We allow intra-team and inter-team interactions, defined by edge $(l, p) \in \mathcal{E}(k)$ with $l, p \in \mathcal{V}_i$, and edge $(l, p) \in \mathcal{E}(k)$ with $l \in \mathcal{V}_i$ and $p \in \mathcal{V}_j$. We define the intra-neighbors as $\mathcal{N}_i^l(k) = \{p\in\mathcal{V}_i | (l, p) \in \mathcal{E}(k)\}$, the inter-neighbors as \mbox{$\mathcal{N}_{-i}^l(k) = \{p\in \bigcup_{i^{\prime}\neq i} \mathcal{V}_{i^{\prime}} | (l, p) \in \mathcal{E}(k)\}$}, and $\mathcal{N}^l(k) = \mathcal{N}_i^l(k) \cup \mathcal{N}_{-i}^l(k)$. 

Each agent $l$ in team $i$ is characterized by a state vector $\bfx_i^l(k)$. The collection of states of team $i$ is given by \mbox{$\bfx_i(k) = [(\bfx_i^1(k))^{\top}, \hdots, (\bfx_i^{\mathsf{M}_i}(k))^{\top}]^{\top}$}, and the collection of states of all the teams except team $i$ is
\[
\bfx_{-i}(k) \kern -0.1cm = \kern -0.1cm [(\bfx_1(k))^{\top}\kern -0.25cm, \hdots, (\bfx_{i-1}(k))^{\top}\kern -0.25cm, (\bfx_{i+1}(k))^{\top}\kern -0.25cm, \hdots, (\bfx_{\mathsf{N}}(k))^{\top}]^{\top}.
\]
Each agent acts in the environment by means of an action vector $\bfu_i^l(k)$. As with $\bfx_i(k)$ and $\bfx_{-i}(k)$, the collection of actions of team $i$ is $\bfu_i(k)$, and the collection of actions of all teams except team $i$ is $\bfu_{-i}(k)$. The dynamics of the game are nonlinear and unknown to the agents. Agents choose their actions based on a policy defined as 
\begin{equation}\label{eq:policy_general}
    \bfu_i^l(k) = \pi_i^l(\bfx^l_{\calN^l}(k), \bftheta_i^l).
\end{equation}
In equation \eqref{eq:policy_general}, $\bfx^l_{\calN^l}(k)$ refers to the state information of the intra- and inter-neighbors of agent $l$, and $\bftheta_i^l$ are parameters that modulate the policy. We consider heterogeneous policies across teammates and teams. The collection of policies of team $i$ is denoted as $\pi_i$, with parameters $\bftheta_i$.

The goal of the teams is to find a sequence of actions that minimize their respective infinite-horizon cost
\begin{equation}\label{eq:cost}
    \min_{\bfu_i(k), 0\leq k<\infty} \sum_{k=0}^{\infty}\gamma c_i(\bfx_i(k), \bfu_i(k), \bfx_{-i}(k), \bfu_{-i}(k)),
\end{equation}
where $0<\gamma<1$ is a discount factor. The \emph{stage} cost of team $i$, $c_i(\bullet)$, is bounded and depends on the states and actions of all the teams of the game at instant $k$. Since we consider an infinite-horizon game, we formulate the task of minimizing \eqref{eq:cost} as finding policies for the agents that follow $\mathcal{G}(k)$.
The definition of the policy in equation \eqref{eq:policy_general} leads to a reformulation of \eqref{eq:cost} in terms of $\pi_i(\bullet)$:
\begin{equation}\label{eq:cost_feedback}
\begin{aligned}
    J^{\infty}_i(&\bfx_i(0), \bfx_{-i}(0), \pi_i(\bullet, \bftheta_i), \{\pi_{j}(\bullet)\}_{j=1, j\neq i}^{\mathsf{N}}) = \\ & \min_{\bftheta_i} \sum_{k=0}^{\infty}c_i(\bfx_i(k), \pi_i(\bullet),\bfx_{-i}(k), \{\pi_{j}(\bullet)\}_{j=1, j\neq i}^{\mathsf{N}}).
\end{aligned}
\end{equation}


\section{Multi-team Distributed Policy Parameterization}\label{sec:distributed_parameterization}
In the absence of a known dynamic model and cost structure, our goal is to learn $\bftheta_i$ $\forall i$ leveraging the only signal available: the cost signal. Therefore, the alternative is to apply a policy gradient method, i.e., use the derivative of $J_i^\infty(\bullet)$ with respect to $\bftheta_i$ to update $\pi_i(\bullet)$. A policy gradient approach for model-free nonlinear multi-agent dynamic games in distributed settings requires a parameterization of $\pi_i(\bullet)$ that supports the graph constraints and is able to integrate information from intra- and inter-neighbors. We propose a parameterization for $\pi_i(\bullet)$ that satisfies this requirement, together with a policy gradient algorithm for learning the policy parameters when only a cost signal and a set of transition tuples $(\bfx_i(k), \bfu_i(k), \bfx_i(k+1))$ is available.

\subsection{Optimal distributed policies in linear quadratic games}

Let $\Pi(\mathbf{x}(k), \bftheta) = [\pi_1(\bullet)^{\top}, \hdots, \pi_{\mathsf{N}}(\bullet)^{\top}]^{\top}$ be the collection of all the team policies, where $\bftheta = [\bftheta_1, \hdots, \bftheta_{\mathsf{N}}]^{\top}$ is the collection of all parameters and $\bfx = [\bfx_1, \hdots, \bfx_{\mathsf{N}}]^{\top}$ is the collection of states of all teams. Model-based infinite games with known linear dynamics, known quadratic cost, and a fixed fully connected communication topology are optimally solved using a linear feedback gain
\begin{equation}\label{eq:optimal_linear_gain}
    \Pi(\mathbf{x}(k), \bftheta^*) = -\mathbf{K}^*\mathbf{x}(k),
\end{equation}
where $\bftheta^* = \mathbf{K}^*$ is the solution of the discrete-time algebraic Ricatti equation of appropriate dimension \cite{kuvcera1972discrete}. The first challenge comes when the system dynamics are unknown and subject to topological constraints that evolve with time.
This is observed, e.g., in \cite{furieri2020learning}, where the model-free learning of the globally optimum linear quadratic regulator in finite-horizon games with time-varying graph constraints is proven.

\begin{theorem}[adapted from \cite{furieri2020learning}]\label{th:furieri}
    Assume unknown linear system dynamics, a quadratic game cost that admits an appropriate compact sub-level set, and a known \textnormal{finite-horizon} sequence of topological constraints given by $\{\mathcal{G}(k)\}_{k=0}^{\mathsf{K}}$ for some finite-time horizon $\mathsf{K}$. Furthermore, assume that the cost gradient is locally dominant, i.e., \mbox{$\mu (J(\Pi(\mathbf{x}(k), \bftheta)) - J^{*}) \leq ||\nabla J(\Pi(\mathbf{x}(k), \bftheta))||_2^2$} for all $\Pi(\bullet)$ and some $\mu > 0$, where $J^*$ is the global minimum of the game and $J(\Pi(\mathbf{x}(k), \bftheta^*))$ is the cost associated to the learned distributed policy. Then, there exist constants \mbox{$\delta>0$} and $\epsilon>0$ such that a distributed policy can be learned, through policy gradient, that satisfies \mbox{$J(\Pi(\mathbf{x}(k), \bftheta^*)) - J^* < \epsilon$} with probability $1 - \delta$.
\end{theorem}

Theorem \ref{th:furieri} sheds light on three key aspects. First, it is possible to learn an optimal distributed policy under time-varying topological constraints in finite-horizon games. In finite-horizon games, the policy in Eq. \eqref{eq:optimal_linear_gain} is replaced by 
\begin{equation}\label{eq:optimal_linear_gain_tv}
    \Pi(\mathbf{x}(k), \bftheta^*) = -\mathbf{K}^*(k)\mathbf{x}(k),
\end{equation}
where the optimal gain is now a sequence of optimal linear feedback gains such that $\bftheta^* = \{\mathbf{K}^*(k)\}_{k=0}^{\mathsf{K}}$ for some finite-time horizon $\mathsf{K}$. Second, in the absence of a known dynamics model, Theorem \ref{th:furieri} requires the evolution of the team graph to be known a priori, which is a limitation in general multi-agent games. Finally, Theorem \ref{th:furieri} assumes linear dynamics with quadratic cost. Overall, Theorem \ref{th:furieri} is a motivating result to seek optimal distributed policies for general multi-agent dynamic games through learning, but key challenges must be addressed to derive a solution for general games.

\subsection{Optimal distributed policies as nonlinear feedback gains}\label{subsec:review}

It is not possible to extend the controller learned using the result in Theorem \ref{th:furieri} to an infinite-horizon problem directly because the approach requires a priori knowledge of the graph connectivity $\{\mathcal{E}(k)\}_k$ and an infinite sequence of control gains $\{\mathbf{K}^*(k)\}_k$, neither of which is possible without additional restrictive assumptions such as periodicity. Therefore, we present two contributions in this regard: (1) our policy parameterization allows time-varying communication in an infinite-horizon problem and (2) extends the formulation in Theorem \ref{th:furieri} to a nonlinear game. 

Specifically, we propose a nonlinear policy of the form
\begin{equation}\label{eq:optimal_gain_ours}
    \Pi(\mathbf{x}(k), \bftheta) = -\mathbf{K}(\mathbf{x}(k), \bftheta)\mathbf{x}(k),
\end{equation}
where $\mathbf{K}(\mathbf{x}(k), \bftheta)$ is a nonlinear function that outputs a feedback gain at each instant and that is parameterized by $\bftheta$. This formulation is a combination of Eqs. \eqref{eq:optimal_linear_gain} and \eqref{eq:optimal_linear_gain_tv}, accounting for the state-dependency and the infinite-horizon nature of the game, since $\mathbf{K}(\mathbf{x}(k), \bftheta)$ can be called infinite times. On the other hand, $\bftheta$ are parameters to be learned from a general cost signal in order to overcome the need of a known (linear) system model and (quadratic) cost.  However, it is yet to be specified how to make $\mathbf{K}(\mathbf{x}(k), \bftheta)$ distributed.

Elaborating on $\mathbf{K}(\mathbf{x}(k), \bftheta)$, Eq. \eqref{eq:optimal_gain_ours} is equivalent to
\begin{equation}\label{eq:extension}
    \Pi(\mathbf{x}(k), \bftheta) \kern -0.1cm = \kern -0.2cm \begin{pmatrix}
        \mathbf{k}^{1,1}_{1,1}(\mathbf{x}(k), \bftheta) & \hdots & 
        \mathbf{k}^{1, \mathsf{M}_{\mathsf{N}}}_{1,\mathsf{N}}(\mathbf{x}(k), \bftheta)
        \\
        \mathbf{k}^{2, 1}_{1,1}(\mathbf{x}(k), \bftheta) & \hdots & 
        \mathbf{k}^{2, \mathsf{M}_{\mathsf{N}}}_{1,\mathsf{N}}(\mathbf{x}(k), \bftheta)
        \\
        \vdots & \vdots & \vdots
        \\
        \mathbf{k}^{\mathsf{M}_{\mathsf{N}}, 1}_{\mathsf{N}, 1}(\mathbf{x}(k), \bftheta) & \hdots & 
        \mathbf{k}^{\mathsf{M}_{\mathsf{N}}, \mathsf{M}_{\mathsf{N}}}_{\mathsf{N}, \mathsf{N}}(\mathbf{x}(k), \bftheta)
    \end{pmatrix}\kern -0.1cm
    \mathbf{x}(k),
\end{equation}
where $\mathbf{k}^{l, p}_{i,j}(\mathbf{x}(k), \bftheta)$ is the block-element matrix that relates agent $l$ of team $i$ with agent $p$ of team $j$. To impose the scalability constraints on the policy, we first approximate $\mathbf{k}^{l, p}_{i,j}(\mathbf{x}(k), \bftheta)$ by $\mathbf{k}^{l, p}_{i,j}(\mathbf{x}(k), \bftheta_i^l)$, such that each agent has a different set of parameters associated to its role. Next, we enforce $\mathbf{k}^{l, p}_{i,j}(\mathbf{x}(k), \bftheta_i^l) = \mathbf{0}$ if $(l, p) \notin \mathcal{E}(k)$ to avoid infeasible propagation of information from non-neighboring agents. However, there is still leakage from non-neighboring information in the dependency of $\mathbf{k}^{l, p}_{i,j}(\mathbf{x}(k), \bftheta_i^l)$ on $\mathbf{x}(k)$. To address this, we approximate $\mathbf{k}^{l, p}_{i,j}(\mathbf{x}(k), \bftheta_i^l)$ by $\mathbf{k}^{l, p}_{i,j}(\bfx^l_{\calN^l}(k), \bftheta_i^l)$. Therefore, Eq. \eqref{eq:extension} is approximated such that the action of an agent at time $k$ is given by
\begin{equation}\label{eq:final_action}
  \mathbf{u}_i^l(k) \kern -0.05cm = \kern -0.35cm \sum^{|\calN^l_i(k)|}_{l^{\prime}=1}\kern -0.3cm\mathbf{k}^{l, l^{\prime}}_{i,i}(\bfx^l_{\calN^l}(k), \bftheta_i^l)\mathbf{x}_i^{l^{\prime}} + \kern -0.35cm \sum^{|\calN^l_{-i}(k)|}_{p=1, j=1} \kern -0.3cm\mathbf{k}^{l, p}_{i,j}(\bfx^l_{\calN^l}(k), \bftheta_i^l)\mathbf{x}_j^{p}.    
\end{equation}
Note that the execution of \eqref{eq:final_action} is fully distributed. To parameterize $\mathbf{k}^{l, p}_{i,j}(\bfx^l_{\calN^l}(k), \bftheta_i^l)$, it is important to note that the number of neighbors changes with time. Thus, we propose to model $\mathbf{k}^{l, p}_{i,j}(\bfx^l_{\calN^l}(k), \bftheta_i^l)$ using self-attention networks \cite{vaswani2017attention}. 
Specifically, the implementation of the expression in \eqref{eq:final_action} is as follows.
The proposed architecture is composed by a sequence of layers, indexed using the subscript \mbox{$w=1,\hdots, \mathsf{W}$}, that perform the following operations:
\begin{gather}\begin{aligned}\label{eq:layers}
    &\kern -0.2cm \mathbf{Q}^i_w =  \mathbf{A}_w^{i}{\mathbf{X}}^i_{w-1}, \hbox{ } \mathbf{K}^i_w = \mathbf{B}_w^{i}{\mathbf{X}}^i_{w-1}, \hbox{ } \mathbf{V}^i_w = \mathbf{C}_w^{i}{\mathbf{X}}^i_{w-1}
    \\
    &\kern -0.2cm\mathbf{Y}^i_w = \chi\left(\mathsf{softmax}\left(\beta(\mathbf{Q}^i_w)\beta((\mathbf{K}^i_w)^{\top})\right) \beta(\mathbf{V}^i_w)\right),
    \\
    &\kern -0.2cm\mathbf{X}^i_w = \psi(\mathbf{D}_w^{i}{\mathbf{Y}}_w^i),
\end{aligned}
\end{gather}
where $\beta(\bullet)$, $\chi(\bullet)$, and $\psi(\bullet)$ are nonlinear activation functions. On the other hand, $\bftheta_i =\{\mathbf{A}_w^{i}, \mathbf{B}_w^{i}, \mathbf{C}_w^{i}, \mathbf{D}_w^{i}\}^{\mathsf{W}}_{w=1}$ are matrices to be learned and of fixed size, regardless of the number of neighbors. To match \eqref{eq:final_action} and \eqref{eq:layers}, the dimensions of the matrices of the last layer are such that $\mathbf{X}^i_{\mathsf{W}}$ has $U \cdot X$ rows and $|\calN^l|$ columns, with $U$ and $X$ the dimensions of the action and state space of the agents. This is because each column of $\mathbf{X}^i_{\mathsf{W}}$ corresponds to the block-element $\bfk^{l,p}_{i,j}(\bfx^l_{\calN^l}(k), \bftheta_i^l)$. Therefore, after a reshaping operation, the robot has all the block-elements $\bfk^{l,p}_{i,j}(\bfx^l_{\calN^l}(k), \bftheta_i^l)$ and state information to compute \eqref{eq:final_action}. 

The use of an attention mechanism to learn the nonlinear feedback gains is dictated by the need of handling time-varying teams of agents. Specifically, attention mechanism are favored by graph topologies induced by proximity, as it is common in multi-robot teams, since spatial
proximity correlates with the learned attention weights. In terms of the impact of attention in the stability of the control policies, one can easily enforce additional constraints in $\mathbf{k}^{l, p}_{i,j}(\bfx^l_{\calN^l}(k), \bftheta_i^l)$ to ensure positive definiteness of $\mathbf{K}(\mathbf{x}(k), \bftheta)$, as it is done in \cite{furieri2020learning, sebastian2023physics}. On the other hand, the self-attention mechanism allows parameter sharing across agents from the same team, which further enhances sample efficiency during training. 

The parameters $\bftheta_i$ \mbox{$\forall i \in \{1, \hdots, \mathsf{N}\}$} are unknown. We propose a policy gradient method to learn the optimal parameters $\{\bftheta_i^*\}_{i=1}^{\mathsf{N}}$. In the case of games with known cost and dynamics, one can find a quadratic approximation to compute the discrete-time Ricatti equation. In model-free cases under unknown cost, only the sequence of costs per team $\{J_i^{\infty}(\bullet)\}_{i=1}^{\mathsf{N}}$. Algorithm \ref{al:policy_gradient} describes the policy gradient training procedure to learn optimal parameters $\{\bftheta_i^*\}_{i=1}^{\mathsf{N}}$. At each iteration $t \in \{1,\hdots, \mathsf{T}\}$, the parameters $\bftheta_i^t$ change as a function of the evolution of the cost signal obtained by running game roll-outs. The game is first initialized with random initial conditions $\bfx(0)$. Then, given $\{\bftheta_i^t\}_{i=1}^{\mathsf{N}}$, the game is played, using the policy in \eqref{eq:final_action}. Notably, Algorithm \ref{al:policy_gradient} is distributed itself, as the optimization steps can be distributed across teams.
Furthermore, to favor exploration, the deterministic policy is embedded in an unbiased stochastic structure, leading to a policy $\rho^l_i(\bfx^l_{\calN^l}(k), \bftheta_i^l)$ such that
\begin{align}
     \mathbb{E}[\rho^l_i&(\bfx^l_{\calN^l}(k), \bftheta_i^l)] = \pi^l_i(\rho_i(\bfx^l_{\calN^l}(k), \bftheta_i^l)), \notag \\
     &\mathbb{E}[\rho_i^l(\bullet) \rho^l_i(\bullet)^{\top}] = \psi^l_i(\bfx^l_{\calN^l}(k), \bfalpha_i^l),    \label{eq:stochastic_regulator}
\end{align}
with $\mathbb{E}[\bullet]$ the expectancy operator and $\psi^l_i(\bullet)$ neural approximators with learned parameters $\bfalpha_i^l$. The actions during training are sampled from the distributions $\rho^l_i(\bullet)$, whereas the deterministic policies $\pi^l_i(\bullet)$ are used at deployment time.
Then, the cost signals $c_i(\bullet)$ are collected, the gradients of the parameters with respect to the cost signals $\partial J^{\infty}_i(\bullet)/ \partial \bftheta_i^t$ are computed, and the parameters are updated using gradient descent with learning rate $\eta > 0$. In this sense, we recall that the cost $J^{\infty}_i(\bullet)$ is assumed bounded, ensuring finite gradients. In practice, we use a multi-agent version of proximal policy optimization (PPO) \cite{schulman2017proximal} that implements Algorithm \ref{al:policy_gradient} and has been proved to be effective in computing the gradients $\partial J^{\infty}_i(\bullet)/ \partial \bftheta_i^t$ and stabilize training, although other centralized, distributed, or value-decomposition multi-agent reinforcement learning algorithms such as QMIX, COMA or DARL1N \cite{foerster2018counterfactual,rashid2020monotonic,wang2022darl1n} can be employed. However, we remark that the focus of the paper is on the policy design, and not on the implementation of the policy gradient algorithm.

\begin{algorithm}[t]
\caption{Policy Gradient with Self-Attention for Model-Free Distributed Nonlinear Multi-Agent Games}\label{al:policy_gradient}
\begin{algorithmic}[1]
\STATE \textbf{Parameters}: initial parameters $\bftheta_i^{0}$ $\forall i \in \{1, \hdots, \mathsf{N}\}$, learning rate $\eta>0$, training steps $\mathsf{T} > 0$, time horizon $\mathsf{K} > 0$
\FOR{iteration $t=1$ to $t=\mathsf{T}$}
    \STATE Roll out game with random initial conditions $\mathbf{x}(0)$ under policies $\rho^l_i(\bullet)$ in \eqref{eq:stochastic_regulator}, parameterized as in \eqref{eq:final_action}
    \STATE Get cost signals $\{c_i(\bullet)\}^{\mathsf{K}}_{k=0}$
    
    \STATE Compute parameter gradients $\partial J^{\infty}_i(\bullet) / \partial \bftheta_i^t$
    
    \STATE Update parameters $\displaystyle{\bftheta_i^{t+1} = \bftheta_i^t - \eta \frac{\partial J^{\infty}_i(\bullet)}{\partial \bftheta_i^t}}$
\ENDFOR 
\STATE Set $\bftheta_i^* = \bftheta_i^{\mathsf{T}}$ $\forall i \in \{1, \hdots, \mathsf{N}\}$
\end{algorithmic}
\end{algorithm}


\section{Results}\label{sec:simulations}
We evaluate our multi-team distributed policy parameterization in different game settings of increasing complexity.
In Sec. \ref{subsec:lqr_results}, we validate the optimality gap of our policy model in a distributed linear quadratic regulation setting, where an optimal LQR baseline exists. Next, in Sec.~\ref{subsec:ilqr_results}, we compare our policy in a nonlinear dynamic multi-agent game against an existing distributed optimal method under known transition and cost models, allowing us to study the optimality gap between model-free and model-based approaches. In Sec. \ref{subsec:marl_results}, we evaluate our policy in a two-team pursuit-evasion game, comparing performance against other existing policy parameterizations.
Finally, Sec. \ref{subsec:robotarium} assesses the effectiveness of our learned policies in a real robot deployment.

\subsection{Distributed linear quadratic regulation}\label{subsec:lqr_results}

The first experiment considers a linear quadratic regulation game under network constraints. We set $\mathsf{N}=5$, $\mathsf{M}_i=1$ for all $i$ and consider unknown linear system dynamics: 
\[
{x}_i(k) = {x}_i(k-1) + b_i{u}_i(k-1) + \omega(k) \quad \forall i \in \mathcal{V}
\]
with ${x}_i(k),{u}_i(k)$ scalars, $\omega(k)$ uniformly distributed in the interval $[-10^{-3}, 10^{-3}]$ as in \cite{furieri2020learning}, $b_i$ uniformly random in $[0, 1]$, and ${x}_i(0)$ uniformly distributed in the range $[-0.1, 0.1]$ for all agents. The cost is \mbox{$J_i^{\infty}(\bullet) =\sum_{k=0}^{\mathsf{K}} \mathbb{E}\big[(\mathbf{x}(k))^{\top}\mathbf{M}(k)\mathbf{x}(k) + (\mathbf{u}(k))^{\top}\mathbf{R}(k)\mathbf{u}(k)\big],$}
with $\mathsf{K}=30$ and $\mathbf{M}(k), \mathbf{R}(k)$ random positive definite matrices with eigenvalues uniformly distributed in the interval $[0.2, 1.0]$. The set of edges $\mathcal{E}(k)$ is randomly generated as a sparse graph for all time steps.

We compare our method against three baselines. The first one computes the centralized linear quadratic regulator (\textsf{LQR}) for finite-horizon games, which is optimal when the network topology is known. The second is the method proposed in \cite{furieri2020learning}, which learns the optimal linear quadratic regulator under known subspace (network) constraints, using a zeroth-order approach. The third one is our method but assuming having access to the graph topology and with $\eta=0.00001$ as learning rate. The fourth method is ours, the same as the previous but does not assume known network topology and instead considers a proximity graph such that \mbox{$(i, j)\in \mathcal{E}(k)$} if and only if $|{x}_i(k) - {x}_j(k)|<0.2$. For the last three methods, we enforce sparsity on the cost by replacing $\mathbf{M}(k)$ by $\mathbf{M}(k)\odot\mathbf{W}(k)$ and $\mathbf{R}(k)$ by $\mathbf{R}(k)\odot\mathbf{W}(k)$, with $\odot$ the element-wise matrix multiplication and $\mathbf{W}(k)$ the adjacency matrix of $\mathcal{G}(k)$.

\begin{figure}[t]
    \centering
    \includegraphics[width=0.8\columnwidth]{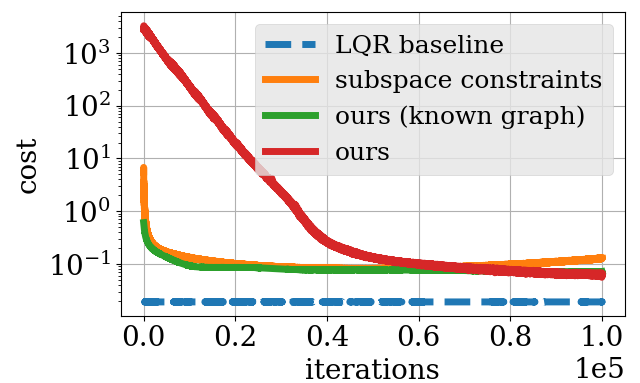}
    \caption{\small Comparison of our method with a centralized \textsf{LQR} baseline, a model-free zeroth-order optimal \textsf{LQR} under subspace constraints \cite{furieri2020learning}, and our method with known graph constraints.}
    \label{fig:furieri_example}
\end{figure}

\begin{figure}[t]
    \centering
    \includegraphics[width=0.75\columnwidth]{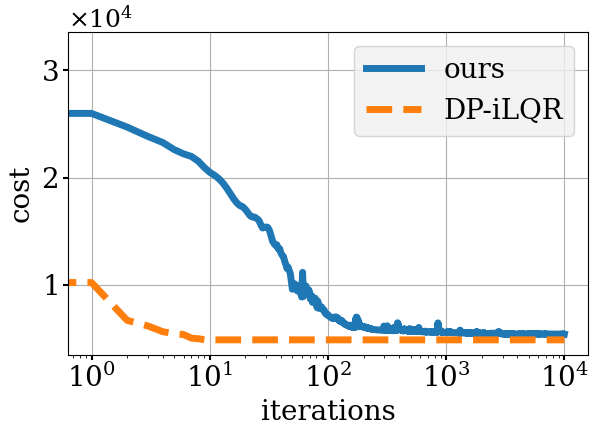}
    \caption{\small Comparison of our method with the Distributed Potential iterative Linear Quadratic Regulator (\textsf{DP-iLQR}) \cite{williams2024potential}.}
    \label{fig:dp_ilqr_example}
\end{figure}

In Fig. \ref{fig:furieri_example}, we can observe that there is an optimality gap between the centralized baseline and the three distributed graph-constrained methods since the pay-off associated to all interactions among the agents is unknown to them. When the graph topology is known at all times, both the method in \cite{furieri2020learning} and ours quickly converge to the optimum under the network constraints, aligned with the theoretical results in \cite{furieri2020learning}. Training of \cite{furieri2020learning} becomes unstable after $60000$ iterations due to the zeroth-order optimization, which uses noisy gradient approximations. Lastly, our policy gradient method with unknown graph topologies is able to convergence to the same cost as the one obtained under the known topology assumption. These results show that there exists an optimality gap between distributed and centralized method and that our method is able to recover the performance under network constraints in multi-agent linear quadratic games.

\subsection{Multi-agent nonlinear games}\label{subsec:ilqr_results}

Next, we evaluate our method in a multi-agent game with nonlinear dynamics and non-quadratic cost. In particular, we reproduce the multi-agent navigation game from \cite{williams2023distributed}, where a set of $\mathsf{N}$ robots navigate in a 2D environment toward assigned locations while avoiding collisions with the other robots. As baseline, we use the Distributed Potential iterative Linear Quadratic Regulator (\textsf{DP-iLQR}) proposed in \cite{williams2024potential}. Given known nonlinear cost and nonlinear multi-agent dynamics, \textsf{DP-iLQR} finds optimal open-loop trajectories for nonlinear multi-agent games that a admit formulation as a potential game.
To match the specifications of the baseline, the game runs for $\mathsf{K}=100$ steps with $\mathsf{M}_i = 1$ \mbox{$\forall i \in \{1, \hdots, \mathsf{N}=7\}$}, and the robots follow unicycle dynamics. The cost $J_i^{\infty}(\bullet)$ is
\[
\begin{aligned}
  J_i^{\infty}(\bullet) =& \sum_{k=0}^{\mathsf{K}} (\mathbf{x}_i(k) - \Bar{\mathbf{x}}_i(k))^{\top}\mathbf{M}_i(k)(\mathbf{x}_i(k)-\Bar{\mathbf{x}}_i(k)) +
  \\
  & \sum_{k=0}^{\mathsf{K}-1}(\mathbf{u}(k))^{\top}\mathbf{R}_i(k)\mathbf{u}(k) + \sum_{j=1 }^{\mathsf{N}}C(d_{i,j}(k)),  
\end{aligned}
\]
with reference states $\{\Bar{\mathbf{x}}_i(k)\}_{k=0}^{\mathsf{K}}$, $d_{i,j}(k) = \|\mathbf{x}_i(k) - \mathbf{x}_j(k)\|$,
\[
C(d_{i,j}(k)) = \begin{cases} \beta( d_{i,j}(k) - d_{\text{prox}}(k))^2 & d_{i,j}(k) < d_{\text{prox}}(k) 
\\ 
0 & \text{otherwise} \end{cases}.
\]
We set $\mathbf{M}_i(k) = \mathbf{I}$, $\mathbf{R}_i(k) = \mathbf{I}$ for all $k$ except $\mathsf{K}$, with $\mathbf{M}_i(\mathsf{K}) = 100\mathbf{I}$. We use Algorithm \ref{al:policy_gradient} with $\eta = 0.001$ and communication radius $r_{\text{comm}} = d_{\text{prox}} = 0.5$ m. 

\begin{figure}[t]
    \centering
    \includegraphics[width=1\columnwidth]{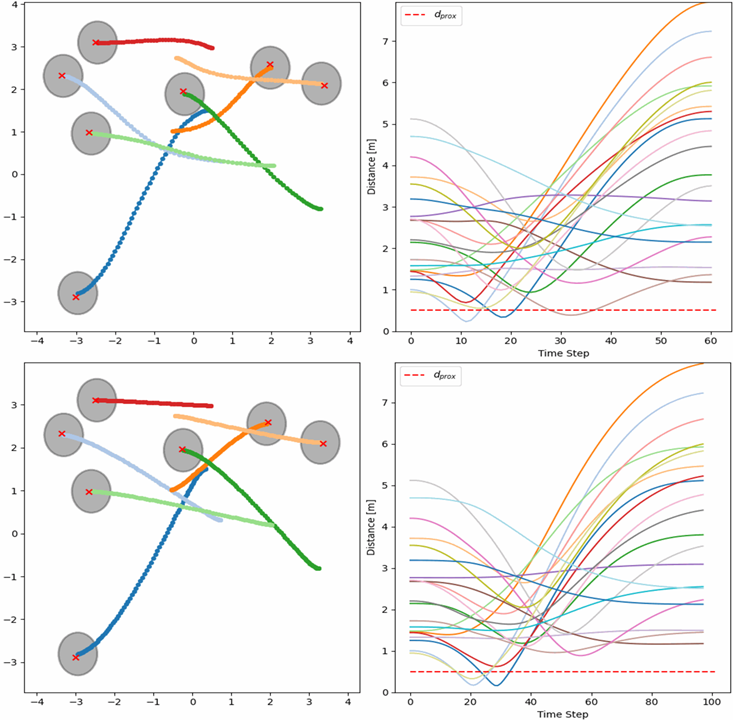}
    \caption{\small Comparison between \textsf{DP-iLQR} at iteration $10$  (\textbf{top row}) and our method at iteration $100$ (\textbf{bottom row}). The agents are shown as grey circles, the goals as red crosses, and the agent trajectories as colored curves in the two pannels. The trajectory evolution (\textbf{left}) and temporal evolution (\textbf{right}) of the agents show that both methods achieve similar position and velocity profiles, as well as security margins with respect to $d_{\text{prox}}$.}
    \label{fig:dp_ilqr_animation}
\end{figure}

\begin{figure}
    \centering
    \begin{tabular}{cc}
    \includegraphics[width=0.42\columnwidth, height=0.38\columnwidth]{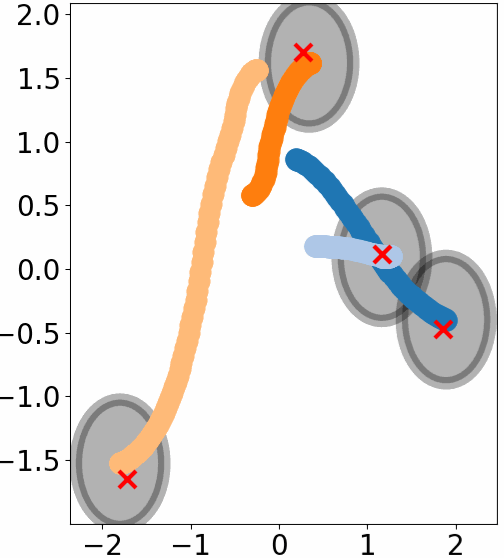}     
         &  
    \includegraphics[width=0.42\columnwidth, height=0.38\columnwidth]{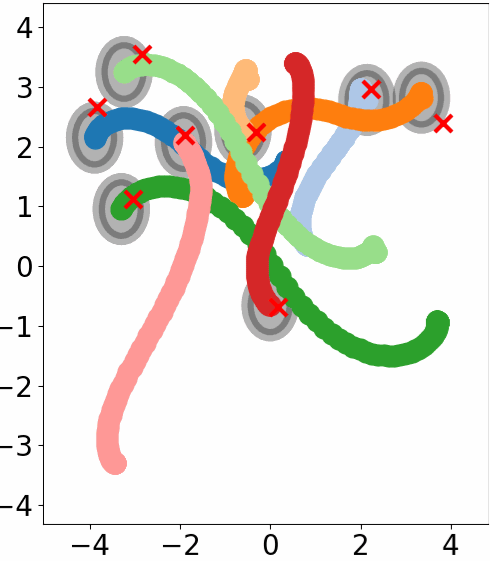}     
         \\
    \includegraphics[width=0.42\columnwidth, height=0.38\columnwidth]{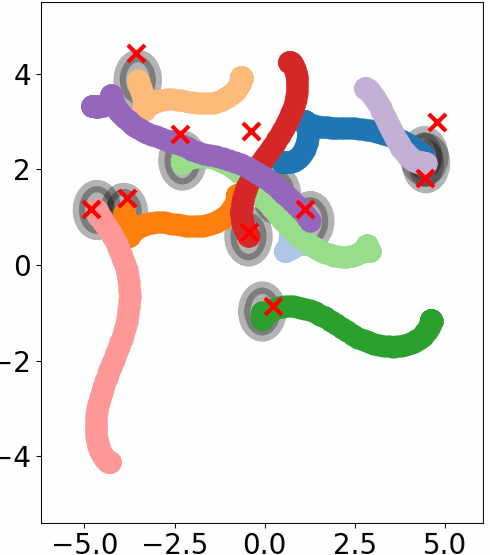}     
         & 
    \includegraphics[width=0.42\columnwidth, height=0.38\columnwidth]{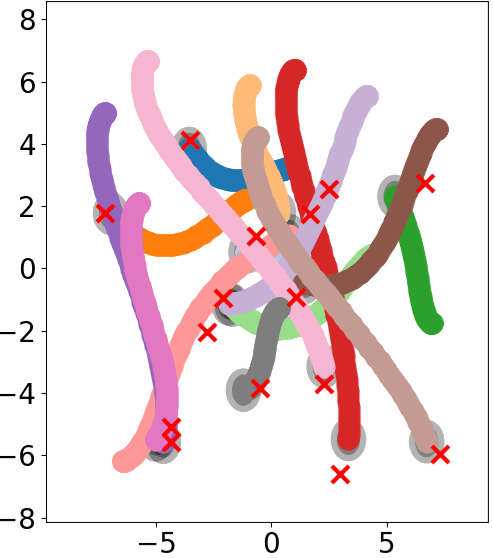}     
    \end{tabular}
    \caption{\small Scalability in the number of agents. Our policy, applied to $\{4, 8, 10, 15\}$ agents, behaves similarly as in the $7$ agents experiment of Fig. \ref{fig:dp_ilqr_example}.}
    \label{fig:scalability}
\end{figure}

As observed in Fig. \ref{fig:dp_ilqr_example}, our policy parameterization achieves near-optimal performance in $100$ training iterations, compared to the $10$ optimization iterations of \textsf{DP-iLQR}. In contrast with the baseline, our solution operates with unknown cost and dynamics, relying only on samples. Qualitatively, in Fig. \ref{fig:dp_ilqr_animation}, we can observe that the differences between the near-optimal performance of our policy and the baseline come from the initial transient behavior. However, in terms of collision avoidance and goal-reaching performance, both methods obtain comparable results, despite the differences in the available information. We also test scalability of our policy across number of agents, $\mathsf{N}=\{4, 8, 10, 15\}$. Fig. \ref{fig:scalability} shows that trajectories arrive to their goals and maintain smoothness for all numbers of agents, with a similar behavior as with $7$ agents in Fig. \ref{fig:dp_ilqr_example}. Collision avoidance constraint violation increases with the number of agents, as density of agents around goals increases. We also run evaluations over $20$ random instantiations of the game, for each number of agents. We compute the mean and standard deviation of $J_i^{\infty}$, normalized by the arena size because the spawning area enlarges with the number of agents to ensure agents do not collide at time 0. As Table \ref{table:scalability} shows, normalized cost remains similar across team sizes. We leave a more systematic scalability analysis as future work. 

\subsection{Pursuit and evasion in BenchMARL}
\label{subsec:marl_results}

Next, we study the performance of our policy model in a pursuit-and-evasion game, Simple Tag, which was introduced in the Multi Particle Environment (MPE) \cite{lowe2017multi}, is implemented in the Vectorized Multi Agent Simulator (VMAS) \cite{bettini2022vmas} and BenchMARL \cite{bettini2024benchmarl}, a multi-agent reinforcement learning suite for training and benchmarking multi-agent policies.
The game evolves in a $2 \times 2$ m$^2$ arena with two circular obstacles of radius $0.2$ m. We consider $\mathsf{N}=2$ teams with $\mathsf{M}_i = 3$ holonomic agents each. 
The agents observe their own position and velocity, the centers of the two obstacles, and the position of all other agents within a radius of $1$ m. The same radius is used for communication with teammates. The agents can also observe the velocities of their neighboring evaders. The evaders (pursuers) receive a positive (negative) reward proportional to their distances from the pursuers (evaders), and are penalized (rewarded) every time they are caught by a pursuer. An episode takes $\mathsf{K}=100$ steps. 

We compare our policy with a multi-layer perceptron (\textsf{MLP}) and a graph neural network (\textsf{GNN}) built as a graph attention network \cite{velivckovic2018graph}. The \textsf{MLP}, with hidden dimensions of $[256, 256, 256]$, is a centralized architecture in the sense that it collects the observations of all agents of all teams as input. The \textsf{GNN} is a distributed architecture that uses only local observations as inputs and also uses hidden dimensions of $[256, 256, 256]$ for the feature encoders. Our policy model uses hidden dimensions of $[64, 64]$ for the attention layers. All networks use hyperbolic tangents as nonlinear activations. To train these policies, we create policy instances for both teams and train them simultaneously using Multi-Agent Proximal Policy Optimization (MAPPO) \cite{yu2022surprising} and Multi-Agent Deep Deterministic Policy Gradient (MADDPG) \cite{lowe2017multi} with the default parameters in BenchMARL for $3\times 10^6$ iterations. We train all policies with two different algorithms to ensure that the results are not specific to a training protocol. 

To assess performance, we conduct a nonparametric binomial paired test  \cite{smeeton2025applied, snyder2025your} as follows. We first set \textsf{Ours} as the candidate architecture for the pursuer team (evader team). Then, we pick either \textsf{Ours}, \textsf{MLP} or \textsf{GNN} as challenged architecture for the pursuer team as well (evader team). With the two challenged policies for the pursuer team, we run randomly initialized $500$ episodes with the same seed against an evader team (pursuer team) using \textsf{MLP} or \textsf{GNN} as baselines. To compare the performance of the two challenged architectures, we consider the cumulative number of catches per episode, \mbox{$\sum_{i=1}^{\mathsf{M}_{\text{eva}}} \left[ \min_{j \in \{1, \ldots, \mathsf{M}_{\text{pur}}\}} \|\bfq^l(k) - \bfq^p(k)\| \le 0.125 \right]$} with $[\bullet]$ denoting the Iverson bracket. A challenged architecture wins in an episode if the pursuers catch more (less) evaders than the other architecture. The significance test, therefore, checks the following null hypothesis: does \textsf{Ours} lose more than the challenged architecture over the baseline? Rejection of the null hypothesis means that there is sufficient evidence of \textsf{Ours} improving upon the challenged architecture.

\begin{table}
    \centering
    \caption{Mean and standard deviation of the total cost $J_i^{\infty}$ over 20 episodes, normalized by area size.}
    \label{table:scalability}
    \begin{tabular}{c|c|c|c|c}
         4 agents
         &  
         7 agents
         &
         8 agents
         &
         10 agents
         &
         15 agents
         \\
         \hline
         277 $\pm$ 116
         &
         241 $\pm$ 103
         &
         308 $\pm$ 134
         &
         248 $\pm$ 84
         &
         290 $\pm$ 73
    \end{tabular}
\end{table}

\begin{figure}[t]
    \centering
    \includegraphics[width=0.9\columnwidth]{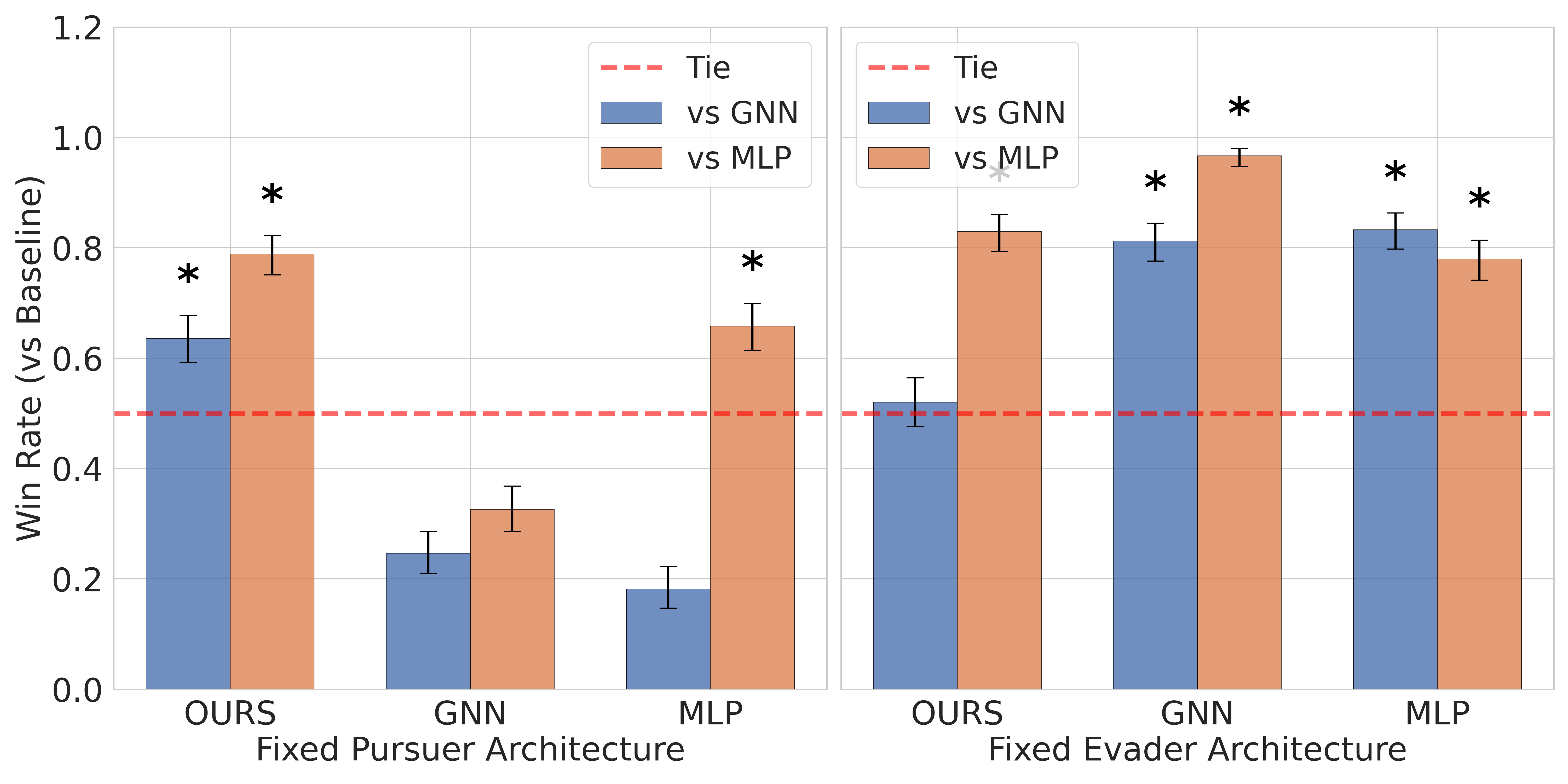}
    \caption{\small Nonparametric binomial paired test over $500$ episodes for policies trained with MAPPO. We report the win rates and Wilson confidence intervals (95\%). Asterisk denotes when the null hypothesis has been rejected with a p-value lower than 0.05.}
    \label{fig:mappo_signifcance_test}
\end{figure}

\begin{figure}[t]
    \centering
    \includegraphics[width=0.9\columnwidth]{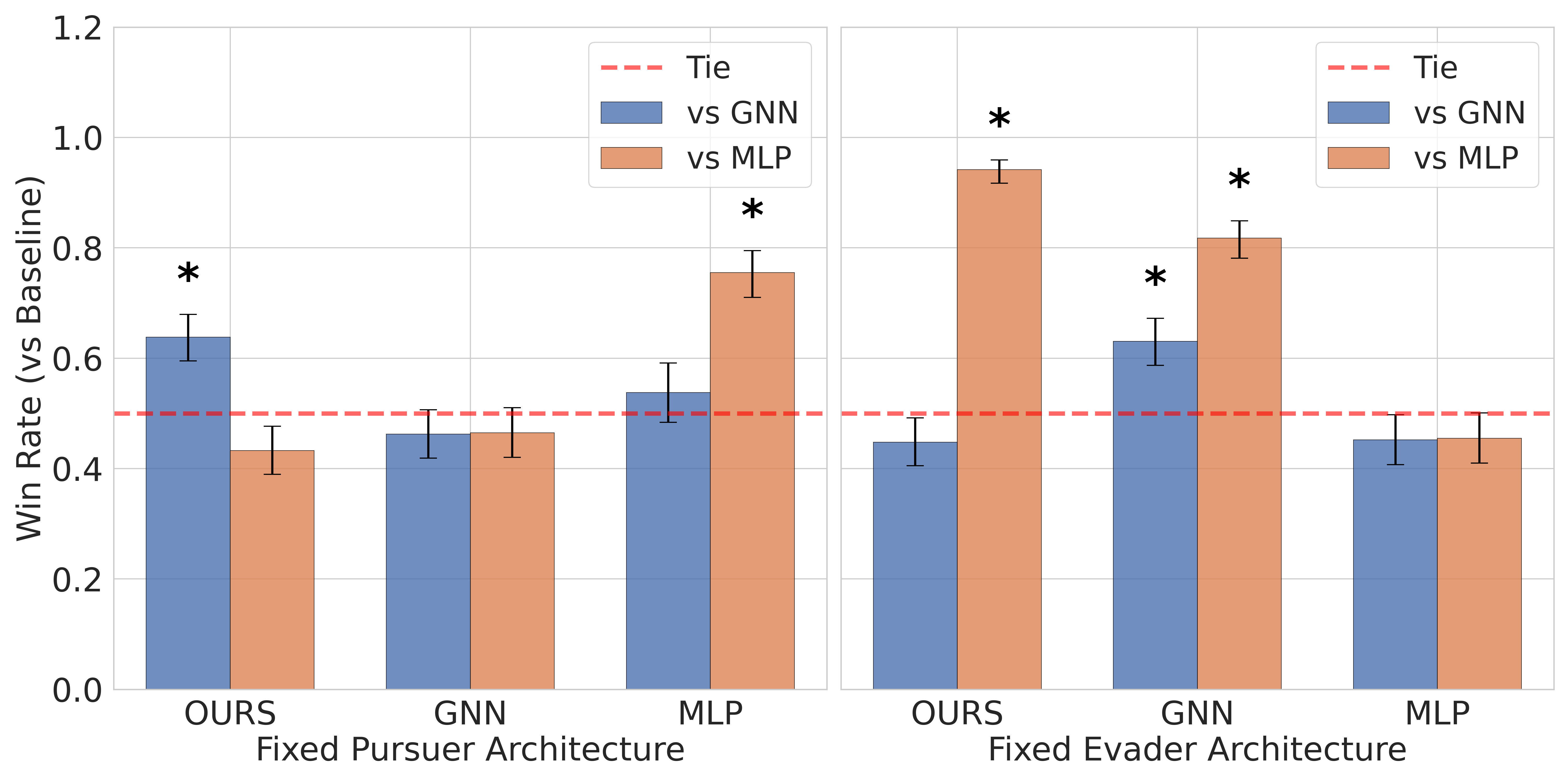}
    \caption{\small Nonparametric binomial paired test over $500$ episodes for policies trained with MADDPG. We report the win rates and Wilson confidence intervals (95\%). Asterisk denotes when the null hypothesis has been rejected with a p-value lower than 0.05.}
    \label{fig:maddpg_signifcance_test}
\end{figure}

As shown in Figures \ref{fig:mappo_signifcance_test} and \ref{fig:maddpg_signifcance_test}, our method is the best model for evaders and pursuers. \textsf{Ours} demonstrates a clear performance edge over \textsf{MLP} and \textsf{GNN}, particularly when trained with MAPPO, where the null hypothesis is rejected 9 out of 12 times. This dominance is most evident in the pursuer's role, where \textsf{Ours} frequently secured win rates exceeding $80\%$. While \textsf{MLP} consistently proved to be the weakest baseline across all tests, the comparison with \textsf{GNN} is more nuanced. \textsf{Ours} generally leads in on-policy MAPPO settings but \textsf{GNN} remains a robust and highly competitive baseline in off-policy MADDPG settings, although with a significantly higher number of parameters. Interestingly, the defensive capabilities of \textsf{Ours} as an evader are more sensitive to opponent strategies. However, even in cases where the null hypothesis cannot be rejected, \textsf{Ours} outperforms (Fig. \ref{fig:maddpg_signifcance_test}, fixed pursuer \textsf{MLP}) or ties (Fig. \ref{fig:mappo_signifcance_test}, fixed evader \textsf{Ours}) versus the \textsf{GNN} baseline. Ultimately, the synergy between our attention-based policy parameterization and MAPPO's centralized advantage estimation provides a superior method for policy learning in multi-team games. In terms of scalability, physical constraints of the scenario limit the number of agents that can be used for evaluation, as can be observed in the qualitative results on our webpage\footnote{\scriptsize \url{https://sites.google.com/view/learning-distributed-games}}. 

\subsection{Pursuit and evasion with real robots}
\label{subsec:robotarium}

Finally, we test our policy parameterization in a real-robot deployment. We zero-shot transfer the policies trained in Sec.~\ref{subsec:marl_results} to the Georgia Tech Robotarium \cite{pickem2017robotarium}, a remotely
accessible, multi-robot research facility that allows for fast integration and deployment of control policies. The arena is $3.2 \times 2.0$ m$^2$ and has a fleet of $20$ non-holonomic robots with radius of $0.11$ m. Due to space constraints, we limit the deployment to $\mathsf{M}_i =3$ agents per team, as in the previous section. Robots include a projection method to match the holonomic commands from the learned policies to the their non-holonomic dynamics, and control barrier functions that ensure safety during robot operation. These are imposed by the Robotarium platform and cannot be modified.

\begin{figure}[t]
    \centering
    \begin{tabular}{cc}
        \fbox{\includegraphics[width=0.4\columnwidth]{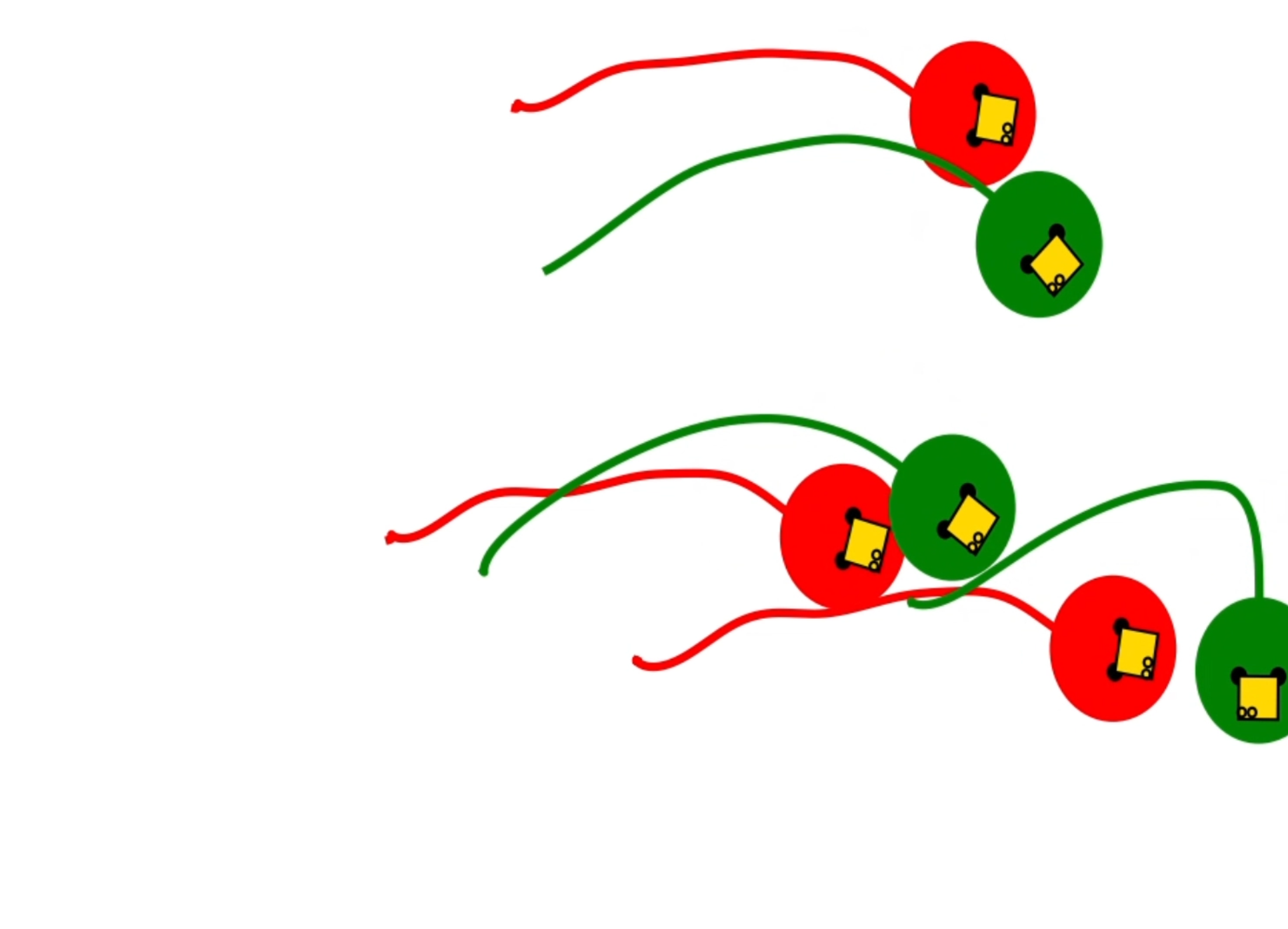}}
        &
        \fbox{\includegraphics[width=0.4\columnwidth]{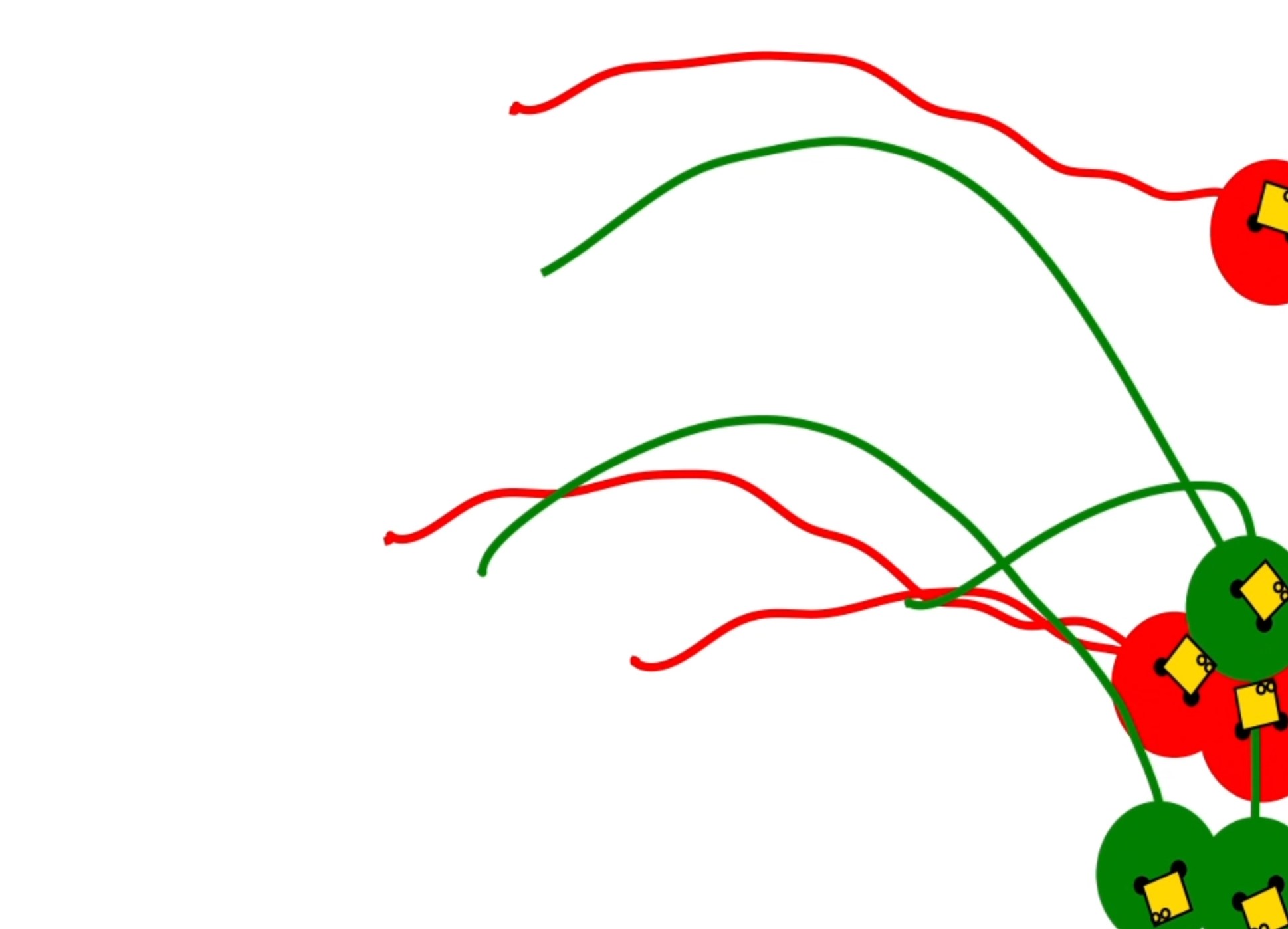}}
    \\
    {\footnotesize (a) close pursuit}
    &
    {\footnotesize (b) pursuers cornering evaders}
    \\
        \fbox{\includegraphics[width=0.4\columnwidth]{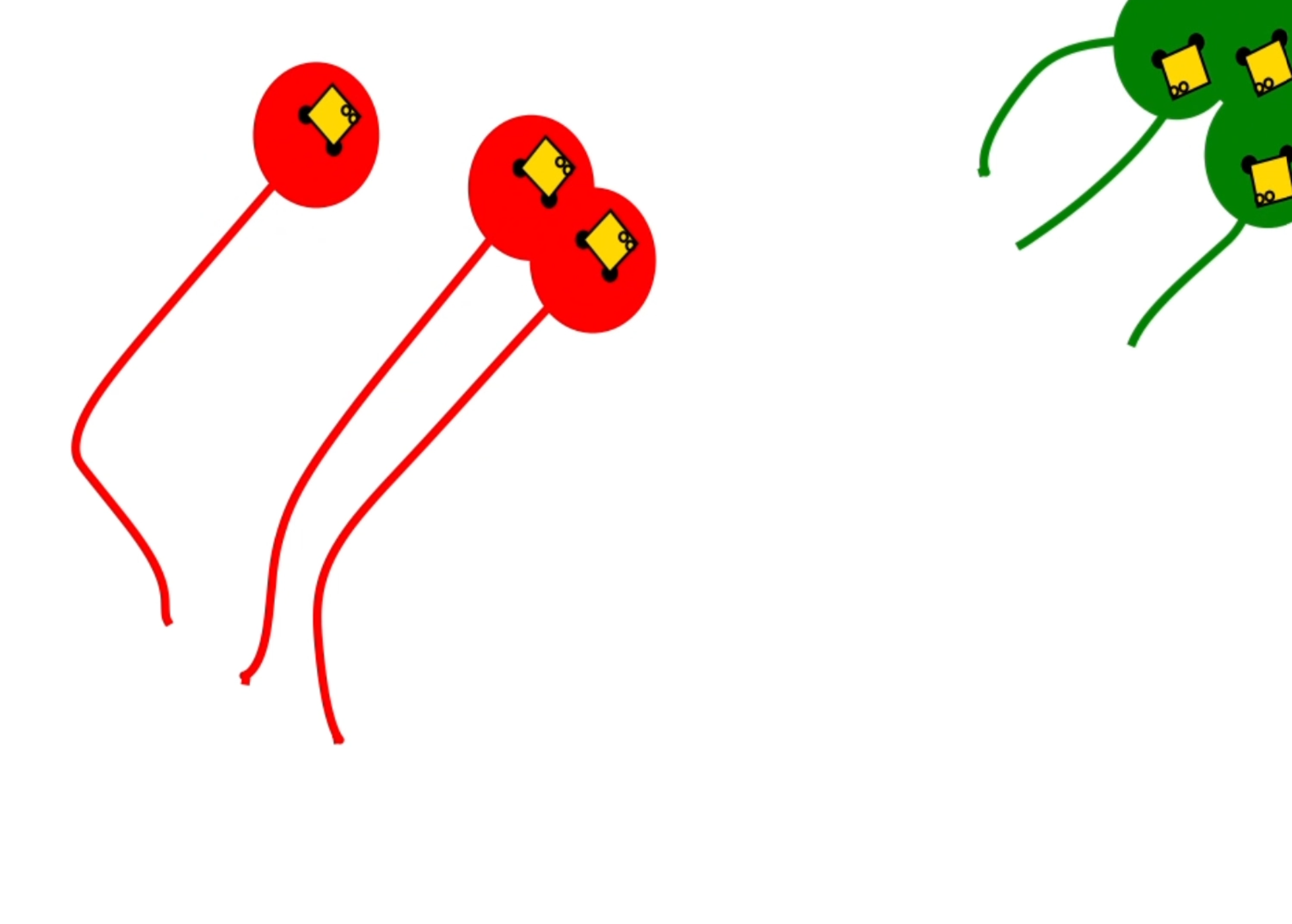}}
        &
        \fbox{\includegraphics[width=0.4\columnwidth]{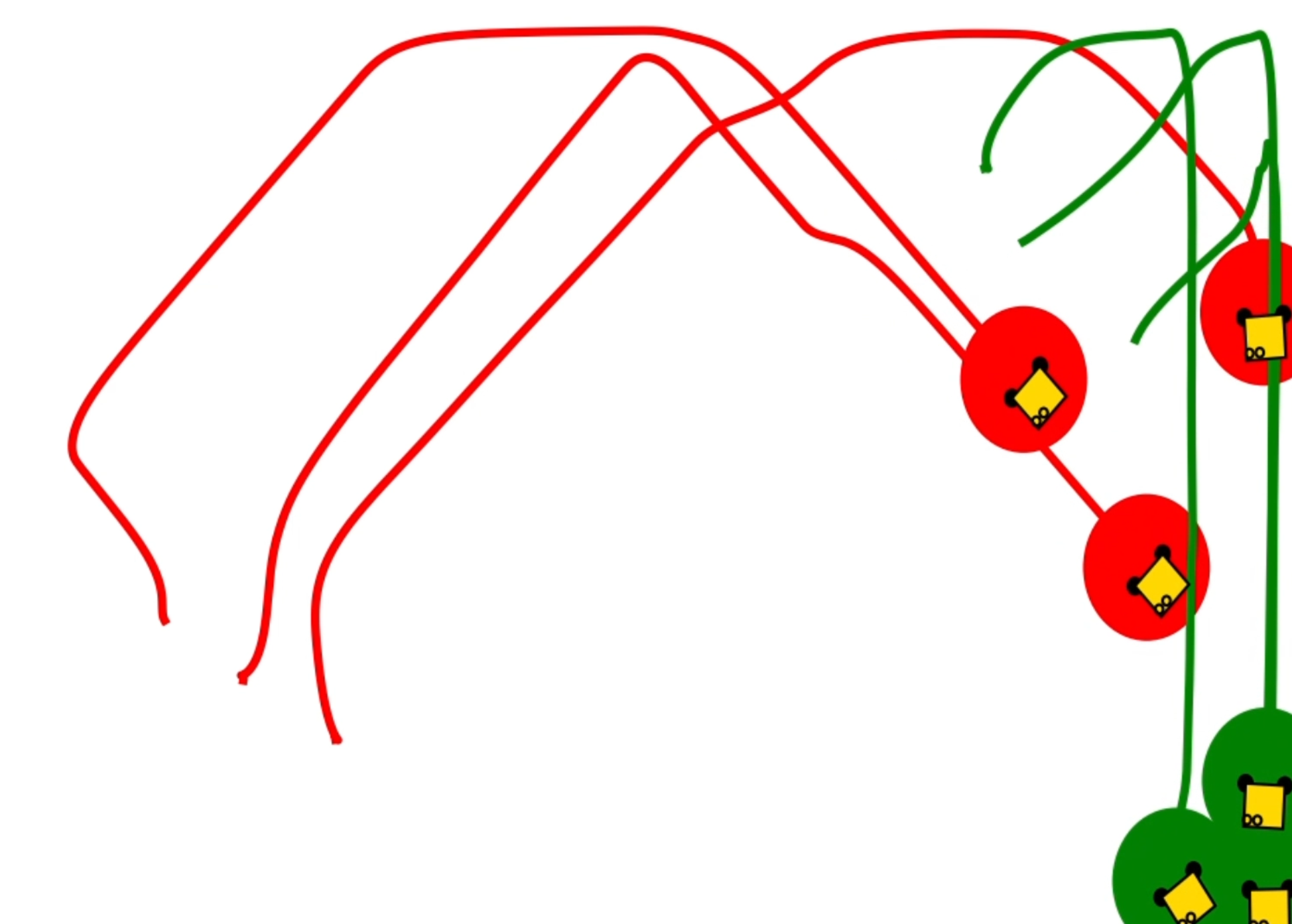}}
         \\
    {\footnotesize (c) evaders sneaking out}
    &
    {\footnotesize (d) pursuers reaction}
    \end{tabular}
    \caption{\small Simulated Robotarium environment modeling the physical interactions and restrictions of the real robots. (top) When both teams are initialized close to each other, the evaders (green) try to escape from the pursuers (red), but the pursuers corner them. Due to the safety filters, the game ends because none of the robots can navigate through others in tight spaces. (bottom) When the teams are initialized far from each other, the evaders go to the most distant corner to minimize the chances of being caught. When the pursuers approach, the evaders sneak out to the other corner, where they finally get trapped due to the safety and space constraints.}\label{fig:simulations}
    ~\\
    \begin{tabular}{cc}
        \fbox{\includegraphics[width=0.4\columnwidth]{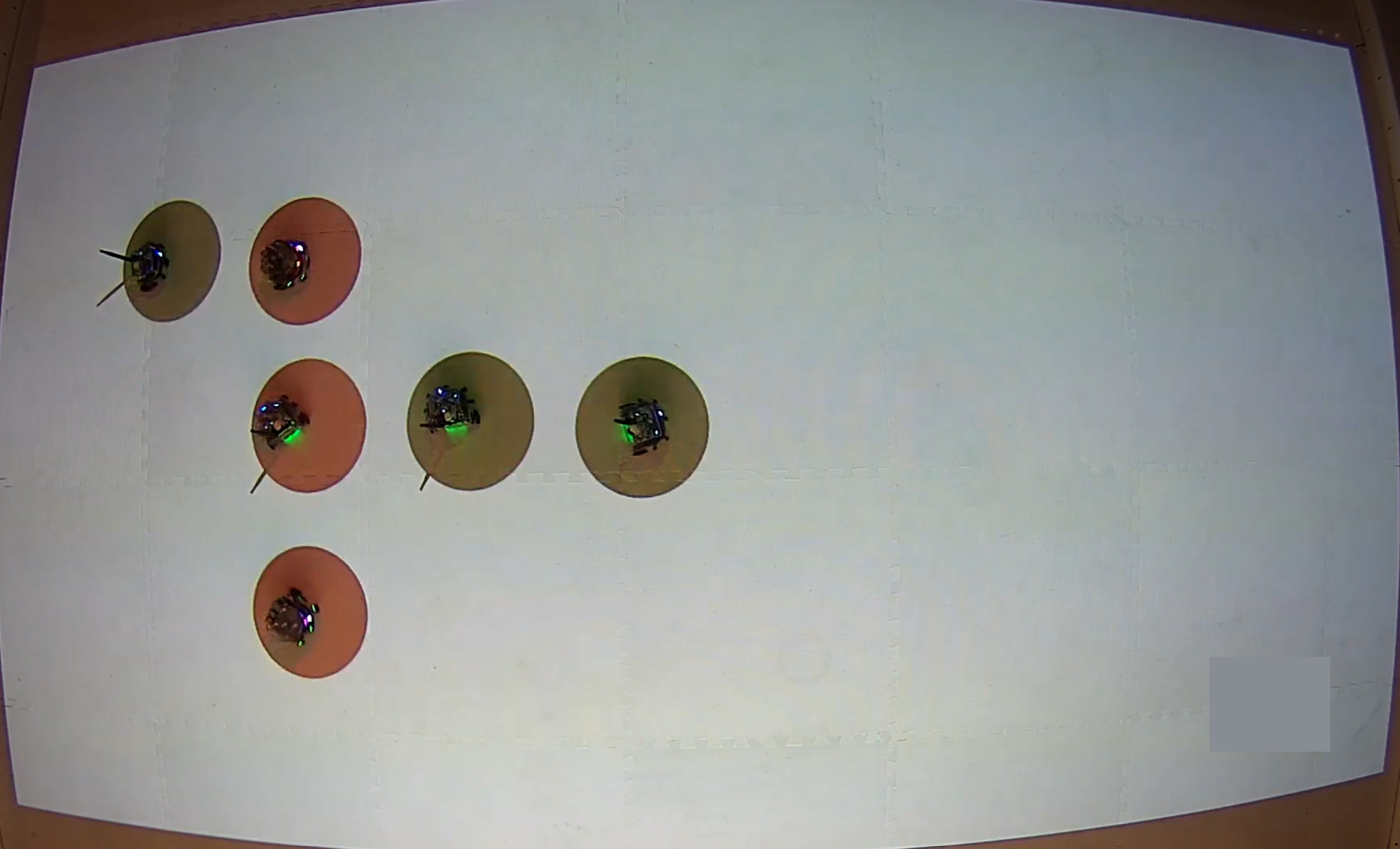}}
        &
        \fbox{\includegraphics[width=0.4\columnwidth]{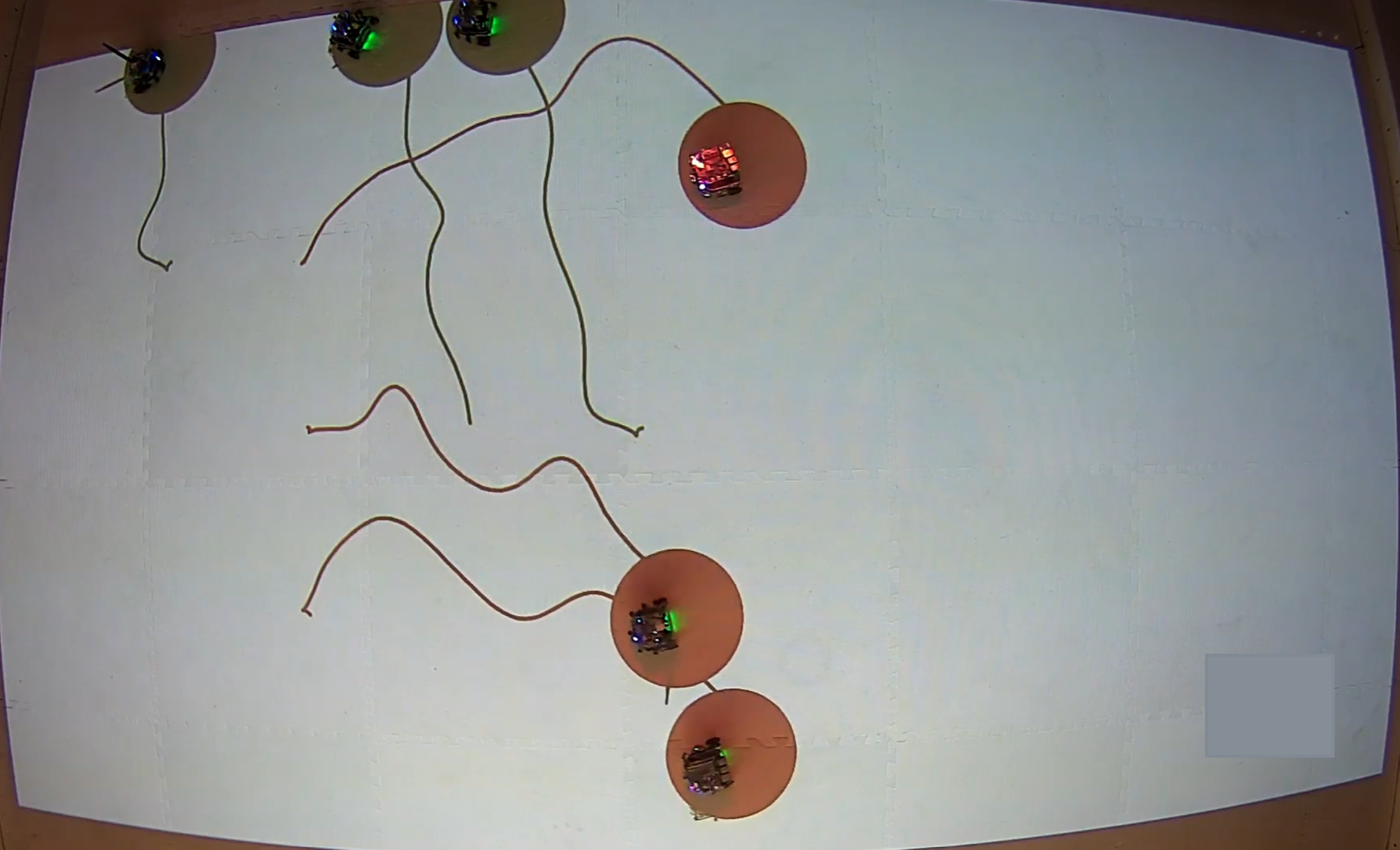}}
         \\
    $0$s
    &
    $12$s
    \\
        \fbox{\includegraphics[width=0.4\columnwidth]{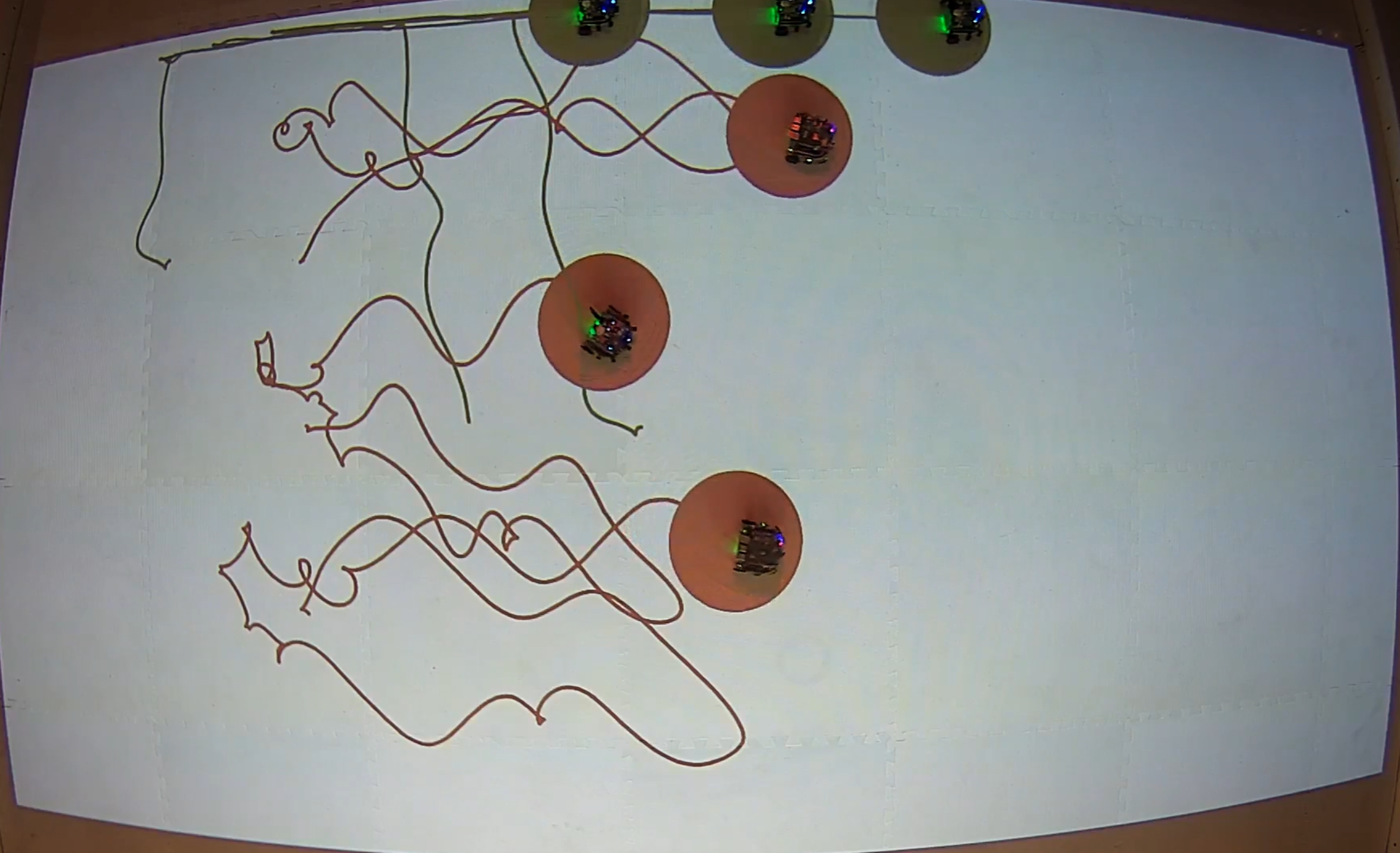}}
        &
        \fbox{\includegraphics[width=0.4\columnwidth]{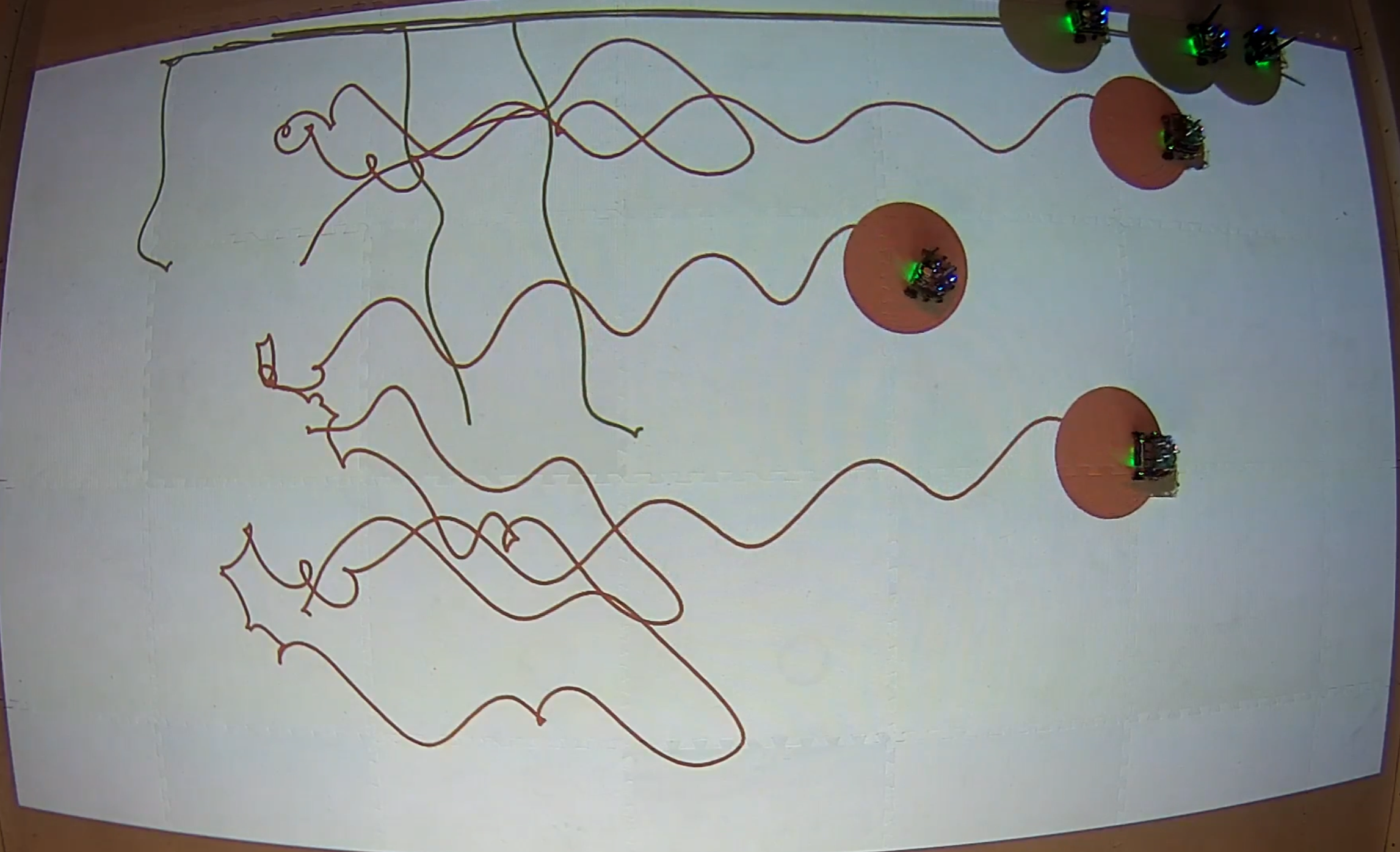}}
    \\
    $40$s
    &
    $52$s
    \end{tabular}
    \caption{\small Real Robotarium deployment. Initially, the evaders (green) can deceive the pursuers (red) and escape to the top-left corner of the arena. However, the pursuers react and try to trap the evaders. The evaders are able to sneak out through the topmost border and escape to the top-right corner being pursued by the other team. The pursuers eventually trap the evaders, finishing the game.}
    \label{fig:experiments}
\end{figure}

Fig.~\ref{fig:simulations} presents snapshots of the Robotarium simulator, which is designed to be as close to the real setting as possible. Despite the gap in realism between the gym-based environment and the real-robot setting, the policies display complex pursuit-and-evasion behaviors. For instance, the evaders deceive the pursuers and react in advance to their efforts to catch them, whereas the pursuers counteract and finally corner the evaders. These complex behaviors also happen in real deployments (Fig. \ref{fig:experiments}), where the pursuers firstly trap the evaders in the top-left corner of the arena but are then tricked by the evaders, who sneak out to the top-right corner of the arena, to finally be trapped again. Videos and further material can be found on the website.

\section{Conclusions}\label{sec:conclusion}
This paper developed a novel method for nonlinear multi-agent dynamic games, focused on model-free multi-team games with time-varying configurations. Inspired by optimal control policies in linear and nonlinear setups, we proposed a novel distributed feedback gain that learns team-level strategies from cost signals. The gains are parameterized with a self-attention structure that handles the dynamic nature of the connectivity of the agents' graph and the diversity in roles/identities of the agents. Experiments showed that our method finds optimal policies under topological constraints without transition and cost models, and that the policies can be deployed in real multi-robot teams. The main limitation of our evaluation is the lack of multi-team robotic benchmarks to comprehensively assess performance across different tasks with different numbers of teams. Existing benchmarks consider either teams of $1$ agent each or only $2$ teams. However, training policies for real applications, like perimeter defense in Fig. \ref{fig:perimeter_defense_example}, require multi-team, multi-agent benchmarks beyond the currently available resources. In future work, it will also be important to investigate the performance of the learned policies under real communication failures and changes in the number of teammates during the game.

\balance
\bibliographystyle{IEEEtran}
\bibliography{IEEEabrv,IEEEexample.bib}

\end{document}